\documentclass[aps,epsfig,twocolumn]{revtex4}
\usepackage{amsmath}
\usepackage{epsfig}
\usepackage{graphics}
\begin{document}
\title{Viscous relaxation and collective oscillations in a trapped Fermi gas near the unitarity limit}

\author{P.\ Massignan}

\affiliation{\O rsted Laboratory, H. C. \O rsted Institute,
Universitetsparken 5, DK-2100 Copenhagen \O, Denmark}

\author{G.\ M.\ Bruun}

\affiliation{Niels Bohr Institute, Blegdamsvej 17, DK-2100
Copenhagen \O, Denmark}

\author{H.\ Smith}

\affiliation{\O rsted Laboratory, H. C. \O rsted Institute,
Universitetsparken 5, DK-2100 Copenhagen \O, Denmark.}

\date{\today{}}

\begin{abstract}
The viscous relaxation time of a trapped two-component gas of fermions
in its normal phase is calculated as a function of temperature and
scattering length, with the collision probability being determined by
an energy-dependent s-wave cross section. The result is used for
calculating the temperature dependence of the frequency and damping of
collective modes studied in recent experiments, starting from the
kinetic equation for the fermion distribution function with mean-field
effects included in the streaming terms.
\end{abstract}

\maketitle

Pacs Numbers: 03.75.Ss, 05.30.Fk, 51.20.+d\

\section{Introduction}

The trapping and cooling of fermions has become one of the central
areas of research within the field of ultra-cold atomic gases.
Such gases offer the exciting prospect of examining the properties
of interacting Fermi gases with unprecedented flexibility. Due to
the existence of Feshbach resonances the interactions between
atoms can be varied almost at will by changing an external
magnetic field, allowing one to study the intriguing problem of a
two-component Fermi gas with a unitarity-limited interaction. It
has been predicted that the gas at low temperatures consists of a
Bose-Einstein condensate (BEC) of tightly bound molecules on the
molecular side of the resonance where $0<k_{\rm F}a\ll 1$, with
$a$ being the scattering length and $k_{\rm F}$ the magnitude of
the Fermi wave vector, and a Bardeen-Cooper-Schrieffer (BCS)
superfluid state on the other side of the resonance where
$0<-k_{\rm F}a\ll1$, with an interesting crossover regime in
between~\cite{Randeria}. Several experimental groups have now
reported clear experimental evidence for a BEC of diatomic
molecules on the molecular side of the resonance from measurements
of the momentum distribution of the gas~\cite{BECmol}. Experiments
have also probed the $a<0$ side of the resonance using a
magnetic-field sweep to the BEC side together with radio frequency
(rf) spectroscopy~\cite{BCSside}.

The study of collective modes is an important spectroscopic tool
for probing the many-body dynamics of atomic
gases~\cite{PethickBook}. Three recent papers report measurements
of the collective modes of a two-component Fermi gas ($^6$Li)
close to the unitarity limit~\cite{Bartenstein,Kinast,Kinast1}.
Although the collective mode spectrum of a normal gas in the
hydrodynamic regime and that of a bulk superfluid gas are
identical~\cite{Baranov}, it was argued in~\cite{Kinast} that the
damping of the modes as a function of temperature should allow one
to distinguish between a normal gas in the hydrodynamic regime and
a superfluid gas.

In the following  we shall examine the collective mode spectrum of
a two-component Fermi gas in its normal phase. Our starting point
is the Boltzmann-Vlasov  equation for the fermion distribution
function $f({\bf r}, {\bf p}, t)$. The effects of atom-atom
interactions enter both the collision integral and the streaming
terms of the equation. Several recent theoretical papers have
considered the effect of collisions on the collective oscillations
of a trapped gas of fermions~\cite{Guery,Vichi,Gehm,Pedri}. Our
approach differs from that of previous authors in that we take
into account both the energy dependence of the scattering cross
section in the collision integral and the effects of the mean
field in the streaming terms. Both these effects are expected to
be important in the region near the resonance. We determine the
appropriate relaxation rate from an approximate solution to the
kinetic equation which is known to give very accurate results for
the viscous relaxation time  in the limits of high and low
temperatures~\cite{HenrikBook}. We then use it to extract the
frequency and damping of the collective oscillations in the
trapped gas.

Since our approach takes full account of the collision processes
and includes the effects of the mean field, it allows for a direct
comparison between the measured and calculated values of the
frequency and damping of collective modes. Our results indicate
that the transverse oscillations measured
experimentally~\cite{Bartenstein,Kinast,Kinast1} are never truly
hydrodynamic in the normal phase, and it identifies the regions at
high and low temperatures in which collisionless behavior is to be
expected. Our calculations suggest that further insight into the
properties of the gas near the unitarity limit may be obtained by
measuring the oscillation frequencies at a fixed magnetic field as
a function of temperature in a regime where the temperature is
comparable to or higher than the Fermi temperature.

The plan of the paper is as follows. In Sec.\ II we introduce the
viscous relaxation time and derive its temperature dependence for
both a uniform and a trapped  gas from approximate solutions to
the kinetic equation for the fermion distribution function. The
kinetic equation also forms the starting point for the calculation
of collective mode frequencies and their damping. In Sec.\ III we
obtain these by taking moments of the equation and present our
results for a trapped gas of $^6$Li-atoms in an axially symmetric
trap. The  effects of interaction in the streaming terms are taken
into account to leading order by adding a mean field to the trap
potential. For realistic scattering lengths and trap parameters
the frequencies of the collective modes are close to their values
in the collisionless limit at high and at low temperatures. In the
intermediate temperature regime the gas approaches hydrodynamic
behavior in the sense that the viscous relaxation rate becomes
comparable to the oscillation frequency. In Sec.\ IV we compare
the calculated frequencies and attenuation of the collective modes
to experiment, and our main findings are summarized in the
concluding Sec.\ V. A detailed account of the moment method used
for solving the Boltzmann-Vlasov equation is given in the
Appendix.

\section{Viscous relaxation time}

We shall consider a two-component Fermi gas of atoms with mass $m$
in its normal phase. The gas may be uniform or trapped in a
potential $V(\bf r)$. We assume that the dynamics is described by
a semi-classical distribution function $f({\bf{r}},{\bf{p}},t)$
which satisfies the Boltzmann equation
\begin{equation}
\frac{\partial f}{\partial t} +\dot{\bf r}\cdot\frac{\partial
f}{\partial{\bf r}} +\dot{\bf p}\cdot\frac{\partial
f}{\partial{\bf p}}=-I[f]\label{Boltzmann},
\end{equation}
where $I$ is the collision integral. The time-development of $\bf
r$ and $\bf p$ is given by the equations of motion
\begin{equation}
\dot{\bf r}={\bf v}=\frac{\bf p}{m};\;\;\;\dot{\bf
p}=-\frac{\partial V}{\partial \bf r}. \label{eqofmot}
\end{equation}
Since this paper  concerns dynamics for which the two components
of the gas move together, we only need to introduce one
distribution function referring to a definite set of internal
quantum numbers, for brevity denoted by "spin" with the two values
$\sigma=\uparrow,\downarrow$. The distribution function $f$ always
refers to a definite spin value, and we have
$f=f_{\uparrow}=f_{\downarrow}$. Since we consider $s$-wave
scattering, the interaction only involves particles with opposite
spin. In the present section the streaming terms on the left hand
side of (\ref{Boltzmann}) do not contain any effects of the
interaction, but later, in Sec.\ III, we shall add these as a mean
field in the equations of motion (\ref{eqofmot}).

The viscous relaxation time, which plays an important part in the
following, is defined in terms of the viscosity. Let us briefly
recall how one determines the viscosity by linearizing the kinetic
equation in the spatial derivatives of the flow velocity  $\bf
u({\bf r})$. For simplicity we take the direction of the flow
velocity as our $x$-axis and assume that it varies in the
$y$-direction, $ {\bf u}=(u_x(y),0,0).$ To calculate the viscosity
we insert a local equilibrium distribution $f_{\rm loc}$ given by
\begin{equation}
f_{\rm loc}({\bf p})=f^0(\epsilon-{\bf u}\cdot{\bf p}), \label{}
\end{equation}
where $\epsilon=p^2/2m$ and $f^0$ denotes the equilibrium Fermi
function, and linearize in the gradient $\partial u_x/\partial y$.
Under stationary conditions (\ref{Boltzmann}) then becomes
\begin{equation}
-\frac{\partial u_x}{\partial y}v_yp_x\frac{\partial f^0}{\partial
\epsilon}=-I[f]. \label{linB}
\end{equation}
The shear viscosity $\eta$ relates the momentum current density
$\Pi_{xy}$, given by
\begin{equation}
\Pi_{xy}=2\int\frac{d^3p}{(2\pi\hbar)^3}v_yp_xf ,\label{momcur}
\end{equation}
to the gradient of the flow velocity according to
$\Pi_{xy}=-\eta\partial u_x/\partial y$. The factor of two
appearing on the right hand side of (\ref{momcur}) arises from
summing over the contributions of the two components ($\uparrow$
and $\downarrow$). In order to establish the concept of a viscous
relaxation time let us make a relaxation time approximation to the
collision integral with a relaxation time $\tau_{\eta}$, which so
far is an unknown quantity,
\begin{equation}
I[f]\approx\frac{f-f^0}{\tau_{\eta}}.
 \label{viscrel}
\end{equation}
The collision integral (\ref{viscrel}) yields together with
(\ref{linB}) and (\ref{momcur})  the viscosity
\begin{equation}
\eta=2\tau_{\eta} \int
\frac{d^3p}{(2\pi\hbar)^3}v_y^2p_x^2\left(-\frac{\partial
f^0}{\partial \epsilon}\right). \label{}
\end{equation}
The total particle density, $n_{\rm tot}$, of the two components
is
\begin{equation}
n_{\rm tot}= 2\int\frac{d^3 p}{(2\pi\hbar)^3}f^0. \label{}
\end{equation}
The angular integration in momentum space yields a factor of
$1/15$, and the ratio $\eta/n_{\rm tot}$ may then after partial
integration be written in terms of energy integrals as
\begin{equation}
\frac{\eta}{n_{\rm tot}}=
\frac{2}{5}\tau_{\eta}\frac{\int_0^{\infty}\epsilon^{5/2}f^0(1-f^0)d\epsilon}{\int_0^{\infty}\epsilon^{3/2}f^0(1-f^0)d\epsilon}.
\label{reltime}
\end{equation}
Analytic expressions for the integrals occurring in
(\ref{reltime}) may be obtained when $kT$ is either small or large
compared to the Fermi energy $\epsilon_{\rm F}=\hbar^2k_{\rm
F}^2/2m$, where the magnitude of the Fermi wave vector, $k_{\rm
F}$, is given by $k_{\rm F}^3=3\pi^2 n_{\rm tot}$. When the
temperature is much larger than the Fermi temperature $T_{\rm
F}=\epsilon_{\rm F}/k$, the equilibrium distribution function is
$f^0\simeq\exp[(\mu-\epsilon)/kT]$, with $\mu$ being the chemical
potential, and (\ref{reltime})  becomes
\begin{equation}
\frac{\eta}{n_{\rm tot}}=kT\tau_{\eta}. \label{highT}
\end{equation}
At low temperatures ($T\ll T_{\rm F}$) we obtain from
(\ref{reltime}) that
\begin{equation}
\frac{\eta}{n_{\rm tot}}=\frac{1}{5}mv_{\rm F}^2\tau_{\eta},
\label{lowT}
\end{equation}
where $v_{\rm F}=\hbar k_{\rm F}/m$ is the Fermi velocity. We
shall use (\ref{reltime}) (and the limiting forms (\ref{highT})
and (\ref{lowT})) to {\it define} the viscous relaxation time
$\tau_{\eta}$ in terms of the viscosity  calculated by taking the
full collision integral into account. The viscous relaxation time
at low temperatures ($T\ll T_{\rm F}$) would thus be defined in
terms of the calculated low-temperature viscosity $\eta_{\rm low}$
according to $\tau_{\eta}=5\eta_{\rm low}/n_{\rm tot}mv_{\rm
F}^2$.

In the presence of the full collision integral $I[f]$, which is a
functional of the distribution function, we linearize the
Boltzmann equation in terms of a small deviation $\delta f$ from
the equilibrium distribution by writing
$f({\mathbf{r}},{\mathbf{p}},t)=f^0({\mathbf{r}},{\mathbf{p}})+\delta
f({\mathbf{r}},{\mathbf{p}},t)$ with
\begin{equation}
\delta f({\mathbf{r}},{\mathbf{p}},t)=
f^0({\mathbf{r}},{\mathbf{p}})[1-f^0({\mathbf{r}},{\mathbf{p}})]\Phi({\mathbf{r}},{\mathbf{p}},t).
\end{equation}
The linearized collision integral becomes a functional of $\Phi$
given by
\begin{gather}
I[\Phi]=\int\frac{d^3p_1}{(2\pi\hbar)^3}\int
d\Omega\frac{d\sigma}{d\Omega}
|{\mathbf{v}}-{\mathbf{v}}_1|\nonumber\\
[\Phi+\Phi_1-\Phi'-\Phi_1']f^0f^0_1(1-{f^0}')(1-{f^0_1}'),\label{collintegral}
\end{gather}
where $d\sigma/d\Omega$ is the differential cross section and
$\Omega$ is the solid angle for the direction of the relative
outgoing momentum
${\mathbf{p}}_r'=({\mathbf{p}}'-{\mathbf{p}}_1')/2$ with respect
to the relative incoming momentum
${\mathbf{p}}_r=({\mathbf{p}}-{\mathbf{p}}_1)/2$~\cite{PethickBook}.
In the present paper, we take the cross section to be given by the
resonant form
\begin{equation}
\frac{d\sigma}{d\Omega}=\frac{a^2}{1+(p_r/\hbar)^2a^2}.
\label{crosssection}
\end{equation}
In the unitarity limit, where $|a|$ tends toward infinity, the
calculated viscous relaxation rate approaches a finite value that
depends on temperature, since the cross section in this case is
determined by the typical value of the wave number $p_r/\hbar$ for
the relative motion.

It should be noted that we neglect in the cross section
(\ref{crosssection}) any effects of the medium, which can be
significant at very low temperatures~\cite{BruunChris}.

\subsection{Viscosity of a uniform gas}

We shall now  calculate the viscosity of a homogeneous gas
starting from Eqs.\ (\ref{Boltzmann}) and (\ref{collintegral}),
using a variational principle  commonly employed in transport
theory. The Boltzmann equation has the form of a linear,
inhomogeneous integral equation, $X=H\Phi$, where the
inhomogeneous term is $X$, and $H$ is an integral operator with
eigenvalues greater than or equal to zero. The use of the Schwarz
inequality $(U,HU)(\Phi,H\Phi)\geq(U,H\Phi)^2$, where
$(\ldots,\ldots)$ denotes a suitably defined scalar product and
$U$ is a trial function, allows one to put a lower bound on the
viscosity, which is proportional to $(\Phi, X)=(\Phi,H\Phi)$. In
the present case we choose $X$  to be equal to $v_yp_x$ and define
the scalar product according to
\begin{equation} (A,B)=\int\frac{d^3
p}{(2\pi\hbar)^3}A({\bf p})B({\bf p})f^0(1-f^0),\label{momcur1}
\end{equation}
where $A$ and $B$ are functions of the momentum. Since the
momentum-space dependence of the left hand side of the Boltzmann
equation involves $v_yp_x$, it is  natural to use a trial function
proportional to $v_yp_x$ as a first approximation. The viscosity
can therefore be approximated by the lower bound
\begin{equation} \eta
=\frac{2}{kT}\frac{(X,X)^2}{(X,HX)},\label{viscexpr}
\end{equation}
where the integral operator $H$ is defined as
 \begin{equation}
H\Phi=\frac{1}{f^0(1-f^0)}I[\Phi].
\end{equation}
The explicit factor of two in the numerator of (\ref{viscexpr})
arises from summing the contributions of the two components. The
corresponding viscous relaxation time is
\begin{equation} \tau_{\eta}
=\frac{(X,X)}{(X,HX)}, \label{taueta}
\end{equation}
which is seen to be independent of the normalization of $X$. The
final expression (\ref{taueta}) is thus an approximate expression,
obtained by a trial function proportional to $v_yp_x$, but it is
known~\cite{HenrikBook} to differ  at high and low temperatures
by only a few per cent from the viscosity obtained from the exact
solution to the Boltzmann equation.

In the classical limit, $T\gg T_{\rm F}$, and for an
energy-dependent scattering cross section given by Eq.\
(\ref{crosssection}) the viscosity, when expressed in terms of the
viscous relaxation time $\tau_{\eta}=\eta/n_{\rm tot}kT$, is (see,
e.g.~\cite{HenrikBook})
\begin{equation}
\frac{1}{\tau_{\eta}}=\frac{8}{5\sqrt{\pi}}n_{\rm
tot}\left(\frac{kT}{m}\right)^{1/2}\bar{\sigma}. \label{Maxwell}
\end{equation}
Here $\bar{\sigma}$ is an effective cross section, which depends
on the ratio $T/T_a$, where the temperature $T_a$ is defined by
\begin{equation}
kT_a=\frac{\hbar^2}{ma^2}.
\end{equation}
In general, we have
\begin{equation}
\bar{\sigma}=\frac{4\pi a^2}{3}\int_0^{\infty}dx
x^7e^{-x^2}(1+x^2T/T_a)^{-1}. \label{barcross}
\end{equation}
For $T\ll T_a$ we obtain from (\ref{barcross}) the classical
result $\bar{\sigma}=4\pi a^2$, while in the opposite limit, $T\gg
T_a$, Eq.\ (\ref{barcross}) yields
\begin{equation}\bar{\sigma}=\frac{4\pi a^2T_a}{3T}
=\frac{4\pi}{3}\frac{\hbar^2}{mkT},\label{barcross1}\end{equation}
which is seen to be independent of the scattering length $a$ and,
apart from a numerical constant, equal to the square of the
thermal De Broglie wavelength.

At low temperatures, $T\ll T_{\rm F}$, one expects on general grounds
that $1/\tau_{\eta}\propto T^2$ due to the restrictions on the
available phase space caused by the occupied states, the so-called
Pauli blocking. The magnitude of $1/\tau_{\eta}$ depends on the
dimensionless quantity $\gamma=(k_{\rm F}a)^2=2T_{\rm F}/T_a$.  The
corresponding variational solution to the Landau-Boltzmann equation of
a Fermi liquid (see Ref.\ \cite{HenrikBook}, Sec.\ 6.2.1) yields
\begin{equation}
\frac{1}{\tau_{\eta}}=2\pi\frac{kT^2}{\hbar T_a}F(\gamma),
\label{Landau}
\end{equation}
where the function $F(\gamma)$ is given by the integral
\begin{equation}
F(\gamma)=2\int_0^1dx\frac{x^5}{\sqrt{1-x^2}}\frac{1}{1+\gamma
x^2}, \label{intF}
\end{equation}
the variable $x$ being equal to the sine of half the angle between
the two incoming particle momenta in a collision. The function
$F(\gamma)$ decreases monotonically from its $\gamma=0$ value
$F(0)=16/15$ to its asymptotic expression $F(\gamma)\simeq
4/3\gamma$ for $\gamma \gg 1$.

In the unitarity limit ($|a|\rightarrow\infty$) the viscous
relaxation rate (\ref{Landau}) becomes independent  of the
magnitude of the scattering length, since $F(\gamma)$ in this
limit is proportional to $1/a^2$. In general, the calculated
relaxation rate tends toward a well defined  value which depends
on temperature, when the scattering length tends toward infinity.
The value of $1/\tau_{\eta}$ at unitarity vanishes as $T^2$ at low
temperatures and as $T^{-1/2}$ at high temperatures.

\begin{figure}
\includegraphics[width=\columnwidth,angle=0,clip=]{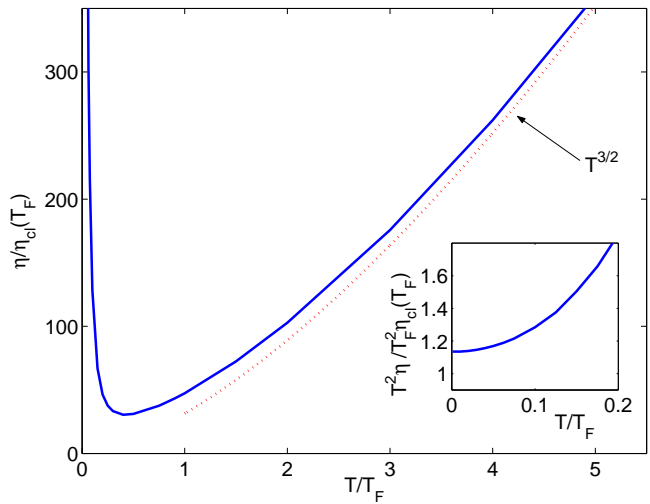}
\caption{The viscosity $\eta$ as a function of
temperature for $k_{\rm F}|a|=4.5$, in units of $\eta_{\rm
cl}(T_{\rm F})=5(mkT_{\rm F}/\pi)^{1/2}/32a^2$, the classical
value of the viscosity for an energy-independent scattering cross
section, evaluated at the Fermi temperature $T=T_{\rm F}$. The
inset illustrates  the low-temperature $T^{-2}$ dependence of the
viscosity.
\label{fig1}
}
\end{figure}

In Fig.~\ref{fig1} we plot the calculated viscosity  as a function of
temperature for the value $k_{\rm F}|a|=4.5$. The inset shows the
viscosity multiplied by $T^2$ in order to illustrate its
characteristic low-temperature behavior given by (\ref{Landau}).
Since $T_a=0.1T_{\rm F}$ for this value of
$k_{\rm F}|a|$, the viscosity at high temperatures is proportional
to $T^{3/2}$. This may be seen by combining the high-temperature
relation $\eta=n_{\rm tot}kT\tau_{\eta}$  with (\ref{Maxwell}) and
(\ref{barcross1}), which  show that $\tau_{\eta}$ is proportional
to $T^{1/2}$ at high temperatures, resulting in $\eta\propto
T^{3/2}$. In the case of energy-independent scattering ($T_a\gg
T_{\rm F}$) the high-temperature viscosity is proportional to
$T^{1/2}$.

\subsection{Viscous relaxation rate of a trapped gas}

In order to apply these results to a trapped atomic cloud we now
include the trap potential in the equilibrium Fermi function. We
consider the harmonic-oscillator potential
\begin{equation}
V({\bf r})=\frac{m}{2}(\omega_x^2x^2+\omega_y^2y^2+\omega_z^2z^2).
\label{osc}
\end{equation}
The average viscous relaxation rate $1/\tau$ is defined by
\begin{equation}
 \frac{1}{\tau}=\frac{\int d^3r(X,HX)}{\int d^3r(X,X)}.
\label{avtaueta}
\end{equation}
Note that the spatial average of (\ref{taueta}) is here carried
out for the denominator and numerator separately. As demonstrated
in the Appendix, this is the quantity that enters as an effective
relaxation rate when we take moments of the kinetic equation in
order to determine the frequency and attenuation  of the
collective modes. The calculation of the average viscous
relaxation rate proceeds as in~\cite{Kavoulakis} (see e.g.\  Eq.\
(38)), where the corresponding rate was obtained for bosons above
the Bose-Einstein condensation temperature, the only modification
being the change of sign in the equilibrium distribution function.
The resulting 5-dimensional integrations were carried out numerically, with varying step
sizes until convergence was achieved.

\begin{figure}
\includegraphics[width=\columnwidth,angle=0,clip=]{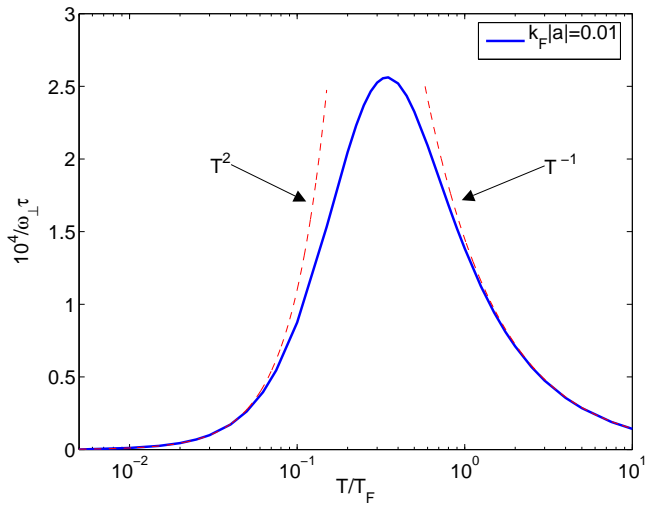}
\caption{The average viscous relaxation rate $1/\tau$ divided by
the transverse trap frequency $\omega_{\perp}$ as a function of
temperature, for $k_{\rm F}|a|=0.01$. The asymptotic temperature
dependencies are indicated by the dashed lines. Note that the
system is highly collisionless, since the maximum value of
$1/\omega_{\perp}\tau$ is about 0.00025.
\label{fig2}
}

\end{figure}

The results shown in Fig.~\ref{fig2} and all following figures
were obtained for a total number $N$ of particles given by
$N=2.8\times 10^5$, which represents a typical value for the
experiments on $^6$Li reported in
\cite{Bartenstein,Kinast,Kinast1}. We use the trap frequencies for
the cigar-shaped cloud of~\cite{Kinast}, i.e.\ an axial frequency
$\omega_z=2\pi\times 70$ Hz and a transverse frequency
$\omega_{\perp}=2\pi\times 1550$ Hz, giving an anisotropy ratio
equal to $ \lambda=\omega_z/\omega_{\perp}=0.045$.

\begin{figure}
\includegraphics[width=\columnwidth,angle=0,clip=]{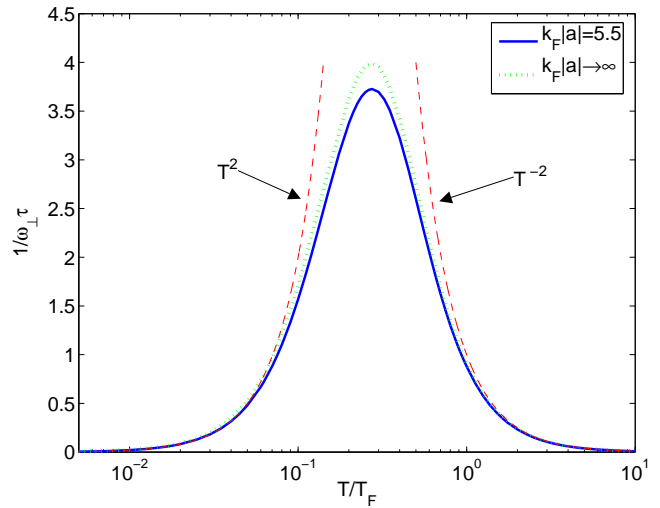}
\caption{The average viscous relaxation rate $1/\tau$ divided by
the transverse trap frequency $\omega_{\perp}$ as a function of
temperature, for $k_{\rm F}|a|=5.5$ corresponding to the
experiment of~\cite{Kinast} at a magnetic field of 870 G. The
asymptotic temperature dependencies are indicated by the dashed
lines. The dotted line is the result obtained in the unitarity
limit $|a|\rightarrow\infty$.
\label{fig3}}
\end{figure}

\begin{figure}
\includegraphics[width=\columnwidth,angle=0,clip=]{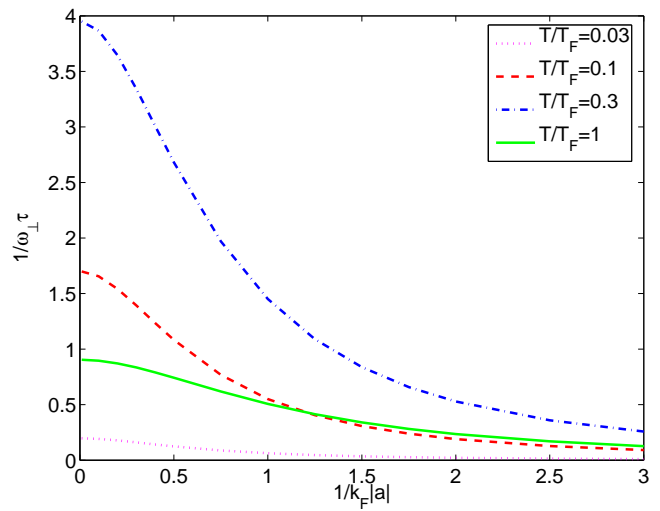}
\caption{The average viscous relaxation rate $1/\tau$ divided by
the transverse trap frequency $\omega_{\perp}$ as a function of
$1/k_{\rm F}|a|$ for four different temperatures. The parameters
$T/T_{\rm F}=0.03$ and $T/T_{\rm F}=0.1$ correspond to the
experimental conditions of Refs.~\cite{Bartenstein}
and~\cite{Kinast1}, respectively.
\label{fig4}
}
\end{figure}

The resulting average viscous relaxation rate is shown in
Figs.~\ref{fig2} and \ref{fig3} for two different values of the
parameter $k_{\rm F}|a|$, one characterizing the regime of weak
coupling and the other the regime near the unitarity limit, where
$k_{\rm F}$ is the magnitude of the Fermi wave vector in the center of
the trap. At low temperatures the relaxation rates are proportional to
$T^2$, and they exhibit in both cases a pronounced maximum at a
temperature somewhat below $T_{\rm F}$. The asymptotic behavior at
high temperatures differs in the two cases. When $k_{\rm F}|a|$ is
much less than unity, the average viscous relaxation rate decreases as
$1/T$ at high temperatures. This may seem at odds with the fact that
for a uniform gas $1/\tau_{\eta}$ according to (\ref{Maxwell}) is
proportional to $T^{1/2}$ in this limit, since $\bar{\sigma}$ is
independent of temperature. However, the average density in a trapped
gas is not a constant, but decreases at high temperatures in
proportion to $T^{-3/2}$, resulting in an average relaxation rate
proportional to $T^{-1}$. When $k_{\rm F}|a|$ is much greater than
unity, the temperature-dependent cross section (\ref{barcross1})
causes the relaxation rate to decrease even more strongly, in
proportion to $T^{-2}$. In the unitarity limit, when $|a|$ approaches
infinity, the average viscous relaxation rate approaches a limiting
value indicated by the dotted curve in Fig.~\ref{fig3}. This is further
illustrated in Fig.~\ref{fig4} where we plot the average viscous relaxation
rate as a function of $1/k_{\rm F}|a|$ for various temperatures.

We have normalized in Figs.~\ref{fig2}, \ref{fig3} and \ref{fig4} the
viscous relaxation rate to the transverse trap frequency
$\omega_{\perp}$ used in the experiments~\cite{Kinast,Kinast1}. The
limiting value of the average viscous relaxation rate for
$|a|\rightarrow\infty$ is seen never to be large compared to
$\omega_{\perp}$, which demonstrates that hydrodynamics cannot be
applied to the transverse motion of the trapped atomic clouds in the
normal phase. In the next section we determine the frequency and
attenuation of the collective modes and obtain results in support of
this general conclusion.

\section{Frequency and attenuation of collective modes}

We proceed now to extract the dispersion relation of the low-lying 
collective modes by solving the linearized Boltzmann-Vlasov
equation (\ref{Boltzmann}) with an appropriate trial function
$\Phi$. The equations of motion (\ref{eqofmot}) are modified to
take into account the mean-field potential $U$ given by
\begin{equation}
U({\bf r})=gn({\bf r})=g[n^0({\bf r})+\delta n({\bf r})],
\end{equation}
where $g=4\pi\hbar^2a/m$ is the interaction constant. The density
$n^0$ denotes the equilibrium density for a single spin, that is
$n^0=n_{\uparrow}^0=n_{\downarrow}^0=n_{\rm tot}^0/2$, and $\delta
n({\bf r})$ similarly denotes the nonequilibrium change in density
for a single spin. The effective potential is thus the sum of $U$
and the harmonic oscillator potential $V({\bf r})$ given by
(\ref{osc}), yielding the equations of motion
\begin{equation}
\dot{\bf r}={\bf v}=\frac{\bf p}{m};\;\;\;\dot{\bf
p}=-\frac{\partial (V+U)}{\partial \bf r}. \label{eqofmot1}
\end{equation}

Let us first establish some useful relations between equilibrium
quantities. We consider the kinetic equation in equilibrium, where
the right hand side of (\ref{Boltzmann}) vanishes, and insert the
equations of motion (\ref{eqofmot1}) which result in
\begin{equation}
\sum_{i=x,y,z}\omega_i\left[p_i\frac{\partial f^0}{\partial\tilde{
r}_i} -\left(\tilde{r}_i+mg\frac{\partial
n^0}{\partial\tilde{r}_i}\right)
        \frac{\partial f^0}{\partial p_i}
\right]=0, \label{eqbe}
\end{equation}
or equivalently
\begin{eqnarray}
\sum_{i=x,y,z}\omega_i\left[p_i\frac{\partial
[f^0(1-f^0)]}{\partial\tilde{r}_i}\right]=\nonumber\\
\sum_{i=x,y,z}\omega_i\left[\left(\tilde{r}_i+mg\frac{\partial
n^0}{\partial \tilde{r}_i}\right)
        \frac{\partial [f^0(1-f^0)]}{\partial p_i}
\right]. \label{equil_BV_eq}
\end{eqnarray}
For convenience we have here introduced the variable
$\tilde{r}_i=m\omega_i r_i$ in terms of which the potential is
$V({\bf r})=\tilde{r}^2/2m$. If we now multiply
(\ref{equil_BV_eq}) by $\tilde{x}p_xp^2_y$ and integrate over both
position and momentum variables,  we obtain
\begin{equation}
\langle p_x^2 p_y^2\rangle-\langle \tilde{x}^2 p_y^2\rangle+ \frac
{m^2} {2}kTg\int d^3 r\, (n^0)^2 =0, \label{momeqbe}
\end{equation}
where $\langle\ldots\rangle$ denotes multiplication by
$f^0(1-f^0)$ and integration over the whole phase space,
\begin{equation}
\langle\ldots\rangle=
 \int d^3 r \int \frac{d^3p}{(2\pi\hbar)^3}\,\ldots f^0(1-f^0). \end{equation}  Using
$f^0(1-f^0)=-(mkT/ p)\partial f^0/\partial p$, we can calculate
analytically the integrals appearing in (\ref{momeqbe}) and obtain
 the virial theorem~\cite{footnote},
\begin{equation}
\label{virial_theorem}
\frac 1 3 E_{\rm kin} - \frac 1 3 E_{\rm
pot} +\frac 1 2 E_{\rm int}=0.
\end{equation}
 Here
\begin{equation}
 E_{\rm kin}=2\int d^3 r
\int\frac{d^3p}{(2\pi\hbar)^3}\frac{p^2}{2m}f^0
\end{equation}
is the kinetic energy,
\begin{equation}
 E_{\rm pot}=2\int
d^3 r\int \frac{d^3p}{(2\pi\hbar)^3}V({\bf r})f^0
\end{equation}
the potential energy, while
\begin{equation} E_{\rm int}=g\int d^3 r (n^0)^2
\end{equation}
is the interaction energy.

 If instead we multiply the equilibrium equation
(\ref{equil_BV_eq}) by the combination $\tilde{x}^2 y p_y$
 and integrate as before, we obtain the equality
\begin{equation}
\langle \tilde{x}^2 p_y^2\rangle - \langle \tilde{x}^2
\tilde{y}^2\rangle - m g \langle \tilde{x}^2 \tilde{y} \frac
{\partial n^0}{\partial \tilde{y}}\rangle=0,\label{useful}
\end{equation}
which will be used in the Appendix to simplify the matrix that we
diagonalize to obtain the frequencies of the collective modes.

To lowest order in the coupling constant $g$, the linearized
version of Eq.~(\ref{Boltzmann}) reads
\begin{gather}
\frac{\partial \Phi}{\partial t}+\sum_{i=x,y,z}\omega_i\left[
{p_i}\frac{\partial\Phi}{\partial\tilde{ r}_i}
-\left(\tilde{r}_i+mg\frac{\partial n^0}{\partial
\tilde{r}_i}\right)
        \frac{\partial\Phi}{\partial p_i}\right.\nonumber\\
\left.-\frac{mg}{f^0(1-f^0)}\frac{\partial \delta n}{\partial
\tilde{r}_i}\frac{\partial f^0}{\partial p_i}
\right]=-\frac{I[\Phi]}{f^0(1-f^0)},\label{lin_Boltzmann}
\end{gather}
where $n^0$ as before denotes the equilibrium  density for a
single spin while the corresponding nonequilibrium change in the
density is
\begin{equation}
\delta n=\int \frac{d^3p}{(2\pi\hbar)^3} f^0(1-f^0)\Phi .
\end{equation}
We shall in the following consider modes for which the drift velocity ${\mathbf{u}}$
has a spatial dependence given by $u_i\propto r_i$. The deviation function $\Phi$
of a fluid moving with velocity ${\mathbf{u}}$ is proportional to ${\mathbf{u}}\cdot{\mathbf{p}}$.
Since acting on ${\mathbf{u}}\cdot{\mathbf{p}}$ with the left side of (\ref{Boltzmann}) generates 
terms like $x^2$, $p_x^2$, etc., we
follow \cite{Usama} in choosing the trial function as
\begin{equation}
\Phi=e^{-i \omega t}\sum_{i=x,y,z}(a_i\tilde r_i^2+b_i \tilde r_i
p_i + c_i p_i^2 ). \label{trialsol}
\end{equation}
We insert this ansatz into the kinetic equation
(\ref{lin_Boltzmann}) and calculate  moments by multiplying with
the product of $f^0(1-f^0)$ and any of the terms
$\tilde{x}^2,\tilde{y}^2,\ldots , p_z^2$ appearing in $\Phi$,
and subsequently  integrating over both $\mathbf r$ and $\mathbf
p$. The result is a homogeneous  set of nine coupled equations for
the nine coefficients $a_x, a_y, \ldots , c_z$ and the frequencies
of the collective modes emerge as the roots of the determinant.
The details of the calculation are given in the Appendix for the
general case when all three trap frequencies are different.

\section{Results and comparison with experiment}

\begin{figure}
\includegraphics[width=\columnwidth,angle=0,clip=]{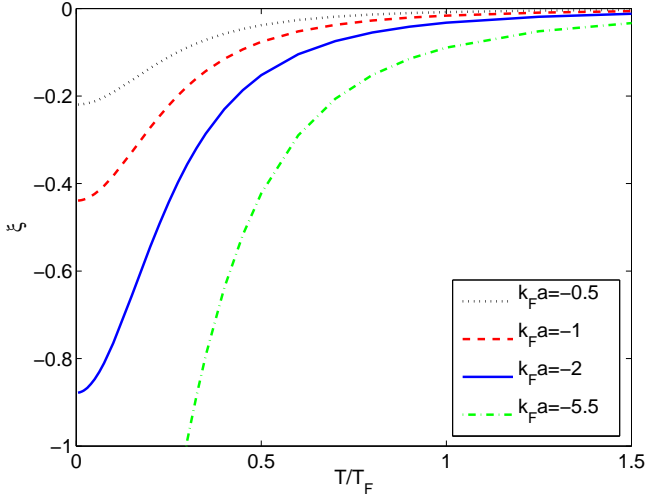}
\caption{The temperature dependence of the parameter $\xi$ given
by (\ref{xi}) for different values of $k_{\rm F}a$. At high
temperatures $|\xi|$ decreases as $T^{-5/2}$.
\label{fig5}
}
\end{figure}

In order to make contact with recent
experiments~\cite{Bartenstein, Kinast, Kinast1} we consider an
axially symmetric trap with $\omega_x=\omega_y=\omega_{\perp}$ and
$\omega_z=\lambda\omega_{\perp}$. We introduce the parameter
\begin{equation}
\xi=\frac{3E_{\rm int}}{2E_{\rm pot}}, \label{xi}
\end{equation}
which, as we shall see, determines the sign and relative magnitude
of the frequency shifts. We shall expand our results to first
order in $\xi$, since our mean-field treatment of the interaction
in the streaming terms of the kinetic equation is only valid when
$|\xi|$ is small compared to unity. The
temperature dependence of $\xi$ is shown in Fig.~\ref{fig5}. In accord
with our first order treatment of the mean field we calculate
$\xi$ by approximating the equilibrium Fermi function, which
enters $E_{\rm int}$ as well as $E_{\rm pot}$, by its value in the
absence of interaction. At high temperatures one finds from
(\ref{xi}) that $|\xi|\propto T^{-5/2}$, since the interaction
energy $E_{\rm int}$ in the classical regime is inversely
proportional to the volume of the cloud ($E_{\rm int} \propto
T^{-3/2}$), while the potential energy is proportional to the
temperature.

The determinant of the matrix, which is derived in the Appendix,
has the form of a polynomial in the frequency $\omega$. The
vanishing of the determinant yields the following equation
\begin{eqnarray} \label{det}
0=\omega \left[ (\omega^2-\omega^2_{\rm hd})\right.\!\!&\!\!-\!\!&\!\!\left.i\omega\tau(\omega^2-\omega^2_{\rm cl})\right]\\
\left[(\omega^2-\omega^2_{\rm hd+})(\omega^2-\omega^2_{\rm hd-})\right.\!\!&\!\!-\!\!&\!\!\left.i\omega\tau(\omega^2-\omega^2_{\rm cl+})
(\omega^2-\omega^2_{\rm cl-})\right],\nonumber
\end{eqnarray}
where $\tau$ is defined by  (\ref{avtaueta}). Note that the
average viscous relaxation rate $1/\tau$ depends on temperature as
illustrated in Figs.~\ref{fig2}, \ref{fig3} and \ref{fig4}.

In general, the solutions to (\ref{det}) have a real and an
imaginary part, $\omega={\rm Re}(\omega) + i {\rm Im} (\omega)$,
which determine the frequency and the damping of the collective
modes, respectively. The (purely real) frequencies appropriate to
the hydrodynamic limit, $\omega\tau\rightarrow 0$, are denoted by
subscript (hd), while those for the collisionless limit,
$\omega\tau\rightarrow \infty$, carry the subscript (cl). To first
order in $\xi$, we find
\begin{gather}
\omega^2_{\rm hd}=2\omega^2_{\perp}\\
 \omega^2_{\rm cl}=4\omega^2_{\perp}(1-\frac{\xi}{2})\\
 \frac{\omega^2_{{\rm
hd}\pm}}{\omega_{\perp}^2}=\frac {5+4\lambda^2\pm\gamma}{3}
                  \pm \xi \frac {4\lambda^4
                  -\lambda^2(5\mp\gamma)+2(5\pm\gamma)}{6\gamma}
\end{gather}
with $\gamma=(25-32\lambda^2+16\lambda^4)^{1/2}$. For $\lambda\ll
1$ the latter are approximately given by
\begin{equation}
\omega^2_{\rm hd+}=\frac{10}{3}\omega_{\perp}^2(1+\frac{\xi}{5})
\label{omegahd+}
\end{equation}
and
\begin{equation}
\omega^2_{\rm hd-}=\frac{12}{5}\omega_{z}^2(1+  \frac{\xi}{20}).
\label{omegahd-}
\end{equation}
The modes labeled + and $-$ are the transverse and axial modes,
respectively, which are studied in the
experiments~\cite{Bartenstein, Kinast, Kinast1}.

 In the
collisionless limit we obtain for elongated traps ($\lambda\ll1$)
\begin{equation}\omega^2_{\rm cl+}=4\omega_{\perp}^2\end{equation}
\begin{equation}\omega^2_{\rm cl-}=4\omega_z^2(1-\frac{\xi}{4})\end{equation}
while for spherical traps ($\lambda = 1$)
\begin{equation}\omega^2_{\rm cl+}=4\omega_{\perp}^2(1+\frac{\xi}{4})\end{equation}
\begin{equation}\omega^2_{\rm cl-}=4\omega_{\perp}^2(1-\frac{\xi}{2}).\end{equation}

\begin{figure}
\includegraphics[width=\columnwidth,angle=0,clip=]{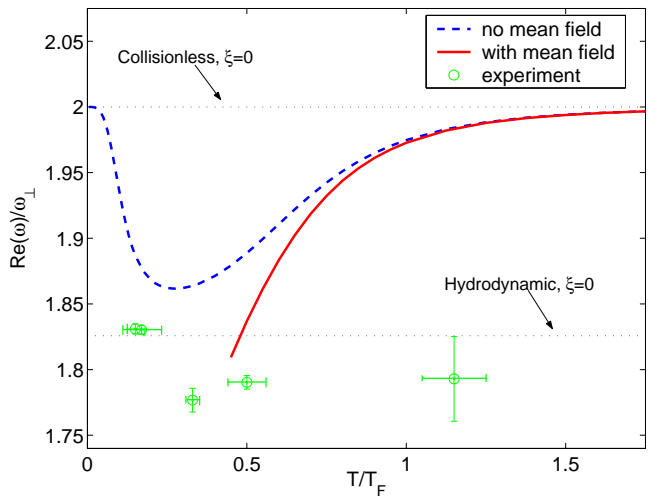}
\caption{The calculated frequency of the transverse (+) mode as a
function of temperature, with and without the mean-field
correction for values of $|\xi|$ less than or equal to 0.5. The
experimental values from~\cite{Kinast} are indicated with the
estimated error bars included. \label{fig6} }
\end{figure}

These results, valid to first order in $\xi$, are in agreement
with those of Pedri \textit{et al.}~\cite{Pedri} when expanded to
first order in $\xi$, but our results differ  to second and higher
order. This is understandable since the form of Eq.\
(\ref{trialsol}) is more general than the scaling ansatz used
in~\cite{Pedri}, which involves six rather than  nine parameters.
However, since our calculation of the frequency shifts caused by
the interaction cannot be trusted beyond first order in $\xi$, our
results are in essential agreement with those of~\cite{Pedri}. Our
work thus extends that of~\cite{Pedri} in the sense that  we
determine $\xi$ and $\tau$ as functions of temperature, thereby
allowing a direct comparison with experiment.

\begin{figure}
\includegraphics[width=\columnwidth,angle=0,clip=]{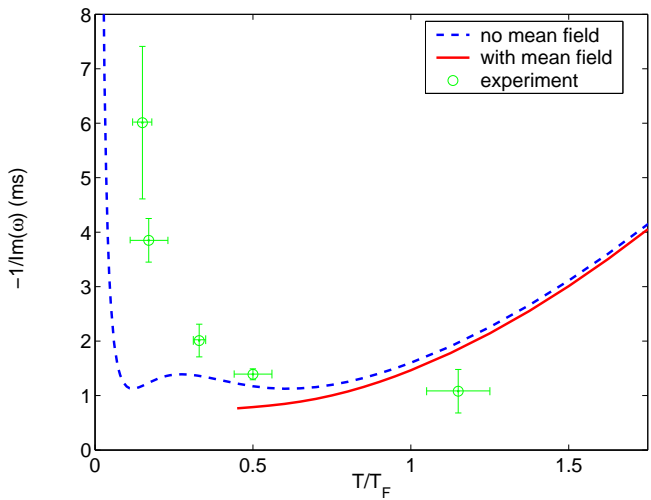}
\caption{The inverse damping rate of the transverse (+)
mode as a function of temperature, with and without the mean-field
correction. The mean-field corrected curve is plotted for values of
$|\xi|$ less than or equal to 0.5. The experimental values
from~\cite{Kinast} are indicated.
\label{fig7}
}
\end{figure}

In Fig.~\ref{fig6} we plot the calculated frequency as a function
of temperature for $k_{\rm F}|a|= 5.5$, which corresponds to the
parameters used in~\cite{Kinast}, along with their experimental
values. Since we assume $|\xi|$ to be small compared to unity, we
show the mean-field curve only in the temperature region where
$|\xi|$ is less than 0.5. There is a clear discrepancy between our
calculated frequency and the experimental result at $T=1.15T_{\rm
F}$ and further work is needed to understand the origin of this.
 The corresponding results for the damping, given by the
imaginary part of the frequency, are shown in Fig.~\ref{fig7}.
In order to compare with the experimental data below $0.5T_{\rm
F}$ in Figs.~\ref{fig6}-\ref{fig7}, it is necessary to improve our treatment of
the interaction effects in the streaming terms of the kinetic
equation.

\begin{figure}
\includegraphics[width=\columnwidth,angle=0,clip=]{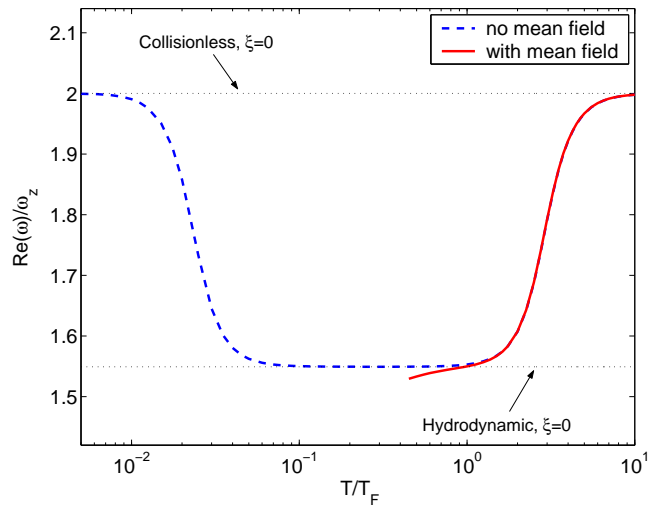}
\caption{The frequency of the axial ($-$) mode as a function of
temperature. \label{fig8} }
\end{figure}

\begin{figure}
\includegraphics[width=\columnwidth,angle=0,clip=]{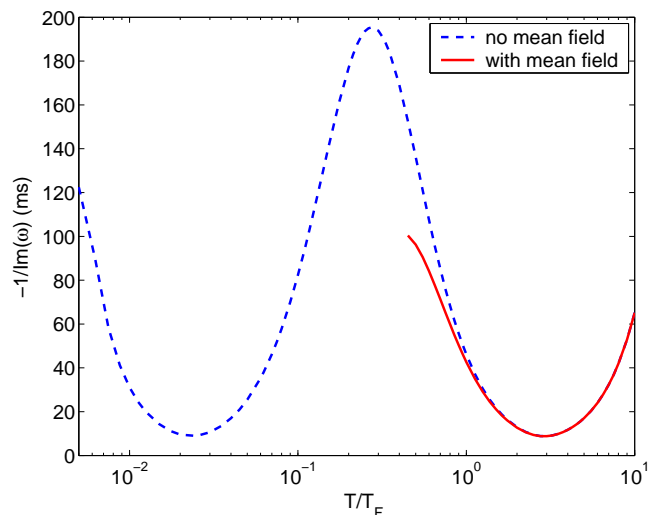}
\caption{The inverse damping rate of the axial ($-$) mode as a
function of temperature. \label{fig9} }
\end{figure}

In Figs.~\ref{fig8} and \ref{fig9} we show results for the real and
imaginary part of the frequency of the axial mode. Since
$\omega_z\ll\omega_\perp$, there is a broad temperature region where
the system behaves hydrodynamically. The damping shows a double-peak
structure that reflects, as temperature is lowered, the transition
between the different regimes, from collisionless to hydrodynamic and
back to collisionless behavior~\cite{Vichi}. The mean-field
corrections in Fig.~\ref{fig8} are seen to be much smaller than those
of Fig.~\ref{fig6}, in agreement with Eqs.~(\ref{omegahd+}) and
(\ref{omegahd-}).

\section{Summary and conclusions}

Starting from a kinetic equation for the semiclassical
distribution function we have calculated  the viscous relaxation
rate which determines the frequency and attenuation of collective modes. By
including interaction effects  as a  mean field in the streaming
terms, we have solved the kinetic equation using a moment method
which takes the conservation laws into account and provides an
accurate account of the damping. Deep in the collisionless regime
the rate of attenuation is proportional to the   viscous
relaxation rate, which  is small compared to the oscillation
frequency. If hydrodynamics is applicable, the viscous relaxation
rate must be much larger than the oscillation frequency, and the
rate of attenuation is then proportional to the oscillation
frequency squared times the viscous relaxation time. Our treatment
in the present work applies to both limits as well as to the
intermediate regime and yields results that allow for a direct
comparison between experiment and theory.

We have demonstrated that for a normal Fermi gas over most of the
temperature range studied
experimentally~\cite{Bartenstein,Kinast,Kinast1}, hydrodynamic theory
does not apply even at the unitarity limit for the transverse
oscillations. It would be interesting to
study further the behavior of the gas very near the unitarity limit, for both
negative and positive $a$, by measuring the oscillation frequencies at
a fixed magnetic field as a function of temperature, thereby testing
the predicted shifts in frequency and attenuation due to the interaction.

\subsubsection*{Acknowledgments}

The authors thank Evgeni Kolomeitsev  and Christopher J.\ Pethick
for useful discussions.

\appendix*
\section{Moments of the Boltzmann-Vlasov equation}

In this appendix we provide details of the calculation of the
$9\times 9$ matrix, from which the frequencies and damping rates can be extracted.

 First we state some useful identities involving the momentum
 variables $p_i$ and the rescaled position variables
 $\tilde{r}_i=m\omega_ir_i$ ($i=x,y,z$),
 \begin{equation}
\langle p_{x}^{2}p_{x}^{2}\rangle=3\langle
p_{x}^{2}p_{y}^{2}\rangle
\end{equation}
together with
 \begin{equation}
\langle p_{x}^{2}p_{y}^{2}\rangle=\frac{1}{3}m^{2}kT E_{\rm
kin}\end{equation} and
 \begin{equation}
\langle \tilde{x}^{2}p_{x}^{2}\rangle=\frac{1}{3}m^{2}kTE_{\rm
pot}.\end{equation} From these it follows that
 \begin{equation}
\frac{\langle \tilde{x}^{2}p_{x}^{2}\rangle}{\langle
p_{x}^{2}p_{y}^{2}\rangle}=\frac{E_{\rm pot}}{E_{\rm
kin}}=\frac{E_{\rm pot}}{E_{\rm pot}-3E_{\rm int}/2}=\frac{1}{1-\xi}.
\end{equation}
 Similar identities hold for the other components of $\mathbf{p}$ and
 $\tilde{\mathbf{r}}$.  They are valid both in the presence and
 absence of the mean field in the equilibrium Fermi function $f^0$.

The following identity holds when the mean field is neglected
in the equilibrium Fermi function,
\begin{gather}
mg\langle \tilde{x}^{2}n^{0}\rangle
=
mg\int\frac{d^{3}p}{(2\pi\hbar)^{3}}\int d^{3}r\ \tilde{x}n^{0}\left(-mkT\frac{\partial f^{0}}{\partial\tilde{x}}\right)
\nonumber
\\
=
m^2kTg\!\int\!\! d^{3}r\
\left[(n^{0})^{2}+\tilde{x}\frac{\partial
n^{0}}{\partial\tilde{x}}n^{0}\right]
=
\frac{m^2}{2}kTE_{\rm
int}
\end{gather}
and implies that
\begin{equation} \frac{mg\langle
\tilde{x}^{2}n^{0}\rangle}{\langle
\tilde{x}^{2}p_{x}^{2}\rangle}\simeq \xi.\end{equation}

In order to obtain the matrix determining the collective modes we
insert (\ref{trialsol}) into (\ref{lin_Boltzmann}) and start
taking moments of (\ref{lin_Boltzmann}) with $\tilde{x}^{2}$,
resulting in
\begin{eqnarray}
&-i\omega[(3a_{1}+a_{2}+a_{3})\langle \tilde{x}^{2}\tilde{y}^{2}\rangle+(c_{1}+c_{2}+c_{3})\langle \tilde{x}^{2}p_{x}^{2}\rangle]+\nonumber\\
&+\omega_{x}b_{1}A_{1}+\omega_{y}b_{2}A_{2}+\omega_{z}b_{3}A_{3}=0.\label{mom_with_rr}\end{eqnarray}
Using the identities given above and Eq.~(\ref{useful}), we find
\begin{eqnarray*}
A_{1} & = & \langle \tilde{x}^{2}p_{x}^{2}\rangle-\langle \tilde{x}^{2}\tilde{x}^{2}\rangle
-mg\langle \tilde{x}^{3}\frac{\partial n^{0}}{\partial\tilde{x}}\rangle=
-2\langle \tilde{x}^{2}p_{x}^{2}\rangle\\
A_{2}&=&\langle \tilde{x}^{2}p_{x}^{2}\rangle-\langle
\tilde{x}^{2}\tilde{y}^{2}\rangle-mg\langle
\tilde{x}^{2}\tilde{y}\frac{\partial
n^{0}}{\partial\tilde{y}}\rangle=0.
\end{eqnarray*}
 The constant $A_{3}$ also vanishes for the same reason as $A_{2}$. Collisions do not 
appear in (\ref{mom_with_rr}) since the combinations 
$\tilde{r}^{2}_i$ (as well as $\tilde{r}_ip_{i}$) are collision invariants, i.e. $I[\tilde{r}^{2}_i]=0.$
We now divide Eq.\ (\ref{mom_with_rr}) by $\langle
\tilde{x}^{2}p_{x}^{2}\rangle$. Since the integral $mg\langle \tilde{y}^{2}\tilde{x}{\partial
n^{0}}/{\partial\tilde{x}}\rangle$ appearing in (\ref{useful}) only introduces frequency shifts of 
second order   in $\xi$, we use here the approximation $\langle
\tilde{x}^{2}\tilde{y}^{2}\rangle/\langle
\tilde{x}^{2}p_{x}^{2}\rangle\simeq 1$.
This yields
\begin{equation}
-i\omega[3a_{1}+a_{2}+a_{3}+c_{1}+c_{2}+c_{3}]-2\omega_{x}b_{1}=0.\label{rr_row}\end{equation}

Next we proceed to take moments with $\tilde{x}p_{x}$ and obtain
\begin{equation}
-i\omega b_{1}\langle\tilde{x}^{2}p_{x}^{2}\rangle
+\omega_{x}[2\langle
\tilde{x}^{2}p_{x}^{2}\rangle(a_{1}-c_{1})+c_{1}m^{2}kTE_{\rm
int}+B]=0,\label{mom_with_rp}\end{equation} where $B$ is given by
\begin{eqnarray*}
B\!&\!=\!&\!-mg\left\langle \frac{\delta n}{f^0(1-f^0)}\left(f^0+\tilde{x}\frac{\partial f^{0}}{\partial\tilde{x}}\right)\right\rangle=\\
\!&\!=\!&\!-mg\langle \tilde{x}^{2}n^{0}\rangle\sum_{i=x,y,z}a_{i}-\frac{m^{2}}{2}kTE_{\rm int}\sum_{i=x,y,z}c_{i}.
\end{eqnarray*}
Dividing Eq.\ (\ref{mom_with_rp}) by $\langle\tilde{x}^{2}p_{x}^{2}\rangle$, we obtain
\begin{equation}
-i\omega
b_{1}+\omega_{x}[(2-\xi)(a_{1}-c_{1})-\xi(a_{2}+c_{2}+a_{3}+c_{3})]=0.\label{rp_row}
\end{equation}

Finally we take moments with $p_{x}^{2}$. Unlike $\tilde{r}_i^2$
and $\tilde{r}_ip_i$ the quantities $p_i^2$ are not separately
collision invariants. By exploiting the fact that the sum
$p_{x}^{2}+p_{y}^{2}+p_{z}^{2}=p^2$ is indeed a collision invariant we
arrive at the equation
\begin{eqnarray}
&-i\omega[(1-\xi)^{-1}(a_{1}+a_{2}+a_{3})+(3+4i/3\omega\tau)c_{1}\nonumber\\
&+(1-2i/3\omega\tau)(c_{2}+c_{3})]+2\omega_{x}b_{1}=0,
\label{pp_row_with_collisions}
\end{eqnarray}
where $\tau$ is given by (\ref{avtaueta}). Eqs.\ (\ref{rr_row}),
(\ref{rp_row}) and (\ref{pp_row_with_collisions}) form
respectively the first, second and third lines of our system of
coupled equations. The other six equations may be obtained from
the three given above by simple permutation of the indices.


\begin{thebibliography}{99}
\bibitem{Randeria} For an account of the early history of the topic, see  M.\ Randeria in
\textit{Bose-Einstein Condensation}, edited by A.\ Griffin, D.\
Snoke, and S.\ Stringari
 (Cambridge University Press, Cambridge 1995).
\bibitem{BECmol}M.\ Greiner, C.\ A.\ Regal, and D.\ S.\ Jin, Nature
\textbf{426}, 537 (2003); M.\ W.\ Zwierlein \textit{et al}.,
Phys.\ Rev.\ Lett.\ \textbf{91}, 250401 (2003); S.\ Jochim
\textit{et al}., Science \textbf{302}, 2101 (2003); T.\ Bourdel
\textit{et al}., Phys.\ Rev.\ Lett.\ \textbf{93}, 050401 (2004).
\bibitem{BCSside} C.\ A.\ Regal, M.\ Greiner, and D.\ S.\ Jin, Phys.\
Rev.\ Lett.\ \textbf{92}, 040403 (2004).
M.\ W.\ Zwierlein
\textit{et al}., Phys.\ Rev.\ Lett.\ \textbf{92}, 120403 (2004);
C.\ Chin \textit{et al}., Science \textbf{305}, 1128 (2004).
\bibitem{PethickBook} See e.g.\ C.\ J.\ Pethick and H.\ Smith,
\textit{Bose-Einstein Condensation in Dilute Gases}
(Cambridge University Press, Cambridge 2002).
\bibitem{Bartenstein} M.\ Bartenstein \textit{et al}., Phys.\ Rev.\ Lett.\ \textbf{92}, 203201 (2004).
\bibitem{Kinast} J.\ Kinast \textit{et al}., Phys.\ Rev.\ Lett.\ \textbf{92}, 150402 (2004).
\bibitem{Kinast1} J.\ Kinast, A.\ Turlapov, and J.\ E.\ Thomas, Phys.\ Rev.\ A \textbf{70}, 051401(R) (2004).
\bibitem{Baranov} M.\ A.\ Baranov and D.\ S.\ Petrov, Phys.\ Rev.\ A \ \textbf{62}, 041601 (2000);
G.\ M.\ Bruun and B.\ R.\ Mottelson, Phys.\ Rev.\ Lett.\ \textbf{87}, 270403 (2001);
G.\ M.\ Bruun and C.\ W.\ Clark, \textit{ibid} \textbf{83}, 5415 (1999).
\bibitem{Guery} D.\ Gu\'{e}ry-Odelin \textit{et al}., Phys.\ Rev.\ A \ \textbf{60}, 4851 (1999).
\bibitem{Vichi} L.\ Vichi, J.\ of Low Temp.\ Phys.\ \textbf{121}, 177 (2000).
\bibitem{Gehm} M.\ E.\ Gehm \textit{et al}.,  Phys.\ Rev.\ A \ \textbf{68}, 011603(R) (2003).
\bibitem{Pedri} P.\ Pedri, D.\ Gu\'{e}ry-Odelin, and S.\ Stringari, Phys.\ Rev.\ A \ \textbf{68}, 043608 (2003).
\bibitem{HenrikBook} H.\ Smith and H.\ H. Jensen, \textit{Transport Phenomena}
(Oxford University Press, Oxford 1989).
\bibitem{BruunChris} G.\ M.\ Bruun and C.\ J.\ Pethick, Phys.\ Rev.\ Lett.\ \textbf{92}, 140404
(2004).
\bibitem{Kavoulakis} G.\ M.\ Kavoulakis, C.\ J.\ Pethick, and H.\ Smith, Phys.\ Rev.\ A
\textbf {61}, 053603  (2000).
\bibitem{footnote} In $d$ dimensions ($d\geq 2$), the virial theorem can straightforwardly be shown to read
$ E_{\rm kin}/d -  E_{\rm pot}/d + E_{\rm int}/2=0$.
\bibitem{Usama} U.\ Al Khawaja, C.\ J.\ Pethick, and H.\ Smith,
J.\ of Low Temp.\ Phys.\ \textbf{118}, 127 (2000).
\end{thebibliography}
\end{document}